\documentclass[prodmode]{acmsmall} 

\usepackage[ruled]{algorithm2e}

\SetAlFnt{\small}
\SetAlCapFnt{\small}
\SetAlCapNameFnt{\small}
\SetAlCapHSkip{0pt}
\IncMargin{-\parindent}

\usepackage{amsmath}
\usepackage[authoryear,round]{natbib}

\acmVolume{X}
\acmNumber{X}
\acmArticle{X}
\acmYear{2016}
\acmMonth{3}

\DeclareMathOperator*{\E}{\mathbb{E}}

\DeclareMathOperator*{\argmin}{arg\,min}

\doi{0000001.0000001}

\issn{}

\begin{document}

\markboth{Fadaei and Bichler}{A Truthful Mechanism for the Generalized Assignment Problem}

\title{A Truthful Mechanism for the Generalized Assignment Problem}
\author{SALMAN FADAEI \\
MARTIN BICHLER}

\begin{abstract}
We propose a truthful-in-expectation, $(1-1/e)$-approximation mechanism for a strategic variant of the generalized assignment problem (GAP).
In GAP, a set of items has to be optimally assigned to a set of bins without exceeding the capacity of any singular bin.
In the strategic variant of the problem we study, values for assigning items to bins are the private information of bidders and the mechanism should provide bidders with incentives to truthfully report their values.
The approximation ratio of the mechanism is a significant improvement over the approximation ratio of the existing truthful mechanism for GAP.

The proposed mechanism comprises a novel convex optimization program as the allocation rule as well as an appropriate payment rule.
To implement the convex program in polynomial time, we propose a fractional local search algorithm which approximates the optimal solution within an arbitrarily small error leading to an approximately truthful-in-expectation mechanism.
The presented algorithm improves upon the existing optimization algorithms for GAP in terms of simplicity and runtime while the approximation ratio closely matches the best approximation ratio given for GAP when all inputs are publicly known.

\end{abstract}

%
%


%
%



\acmformat{Salman Fadaei,
and Martin Bichler, 2016. A Truthful Mechanism for the Generalized Assignment Problem.}

\begin{bottomstuff}

Author's addresses: Salman Fadaei (corresponding author), salman.fadaei@gmail.com. This work was done while the author was a graduate student in Department of Informatics, TU M\"unchen, Munich.
Martin Bichler, Department of Informatics, TU M\"unchen, Munich, Germany. 
\end{bottomstuff}

\maketitle

\section{Introduction}

We analyze the generalized assignment problem (GAP) in an environment where valuations are private information of distributed decision makers.\footnote{A one-page abstract of this paper appeared in WINE 2014.}
In GAP, a set of $m$ items has to be assigned to a set of $n$ bins.
Each bin associates a different value and weight to each item and has a limited capacity.
An allocation may assign each bin a subset of items not exceeding the capacity of the bin.
For each of these subsets, the valuation is additive in the values of items contained in the subset.
The goal is to find a feasible assignment of items to bins to maximize social welfare, the sum of generated values by the assignment.

GAP has also been defined in the literature as a (closely related) minimization problem.
In the minimization GAP, the assignment of items to bins incurs costs; the goal of this optimization problem is to find a feasible assignment of minimum total cost.
From an optimization point of view, 
these two variants of GAP are equivalent \citep{martello1992generalized}.

GAP is a well-known problem in combinatorial optimization and operations research.
It can be considered as a generalization of the problem of finding a maximum weight matching in a weighted bipartite graph, the assignment problem  \citep{kuhn1955hungarian, ferland1998generalized}.
GAP has many real-world applications including applications in resource scheduling problems such as machine scheduling, classroom scheduling and employee scheduling \citep{zimokha88}.
GAP is commonly applied in transportation and routing \citep{ruland1999model,fisher1981generalized}.
There is a long list of reported applications of GAP in telecommunication, production planning and facility location applications \citep{bressoud2003optimal,dobson2001batch,ross1977modeling,klastorin1979note}.
GAP can also be applied in supply chain and logistics.
\citet{kalagnanam2001computational} discuss the computational complexity of clearing markets in a double auction and formulates the problem with GAP.
For a survey of applications of GAP, we refer the reader to \citet{oncan2007survey}.

In many situations, weights and capacities are intrinsic attributes of items and bins and are therefore readily known and verifiable.
For example, in process industries such as paper and steel,  standard geometries such as width, length and weight are used, and buyers typically bid for rolls of paper or steel of a desired width \citep{kalagnanam2001computational}.

By contrast, the associated value of a specific assignment has to be extracted through communication with buyers.
In service areas, for example, the cost of providing a service to a certain group (weight) at an affordable cost (capacity) are known, although the business value of a service is only known to the recipient of the service.

Examples like these motivate the study of a strategic variant of GAP where valuations are assumed to be private information known only to the bidders while weights and capacities are publicly known.
There are two obstacles in finding a solution to this strategic variant.
First, GAP is a NP-hard problem. Hence, computing its optimal social welfare is intractable even if the valuations are known.
Second, maximizing the social welfare necessitates knowing valuations, but this is private information and can be truthfully extracted solely by providing incentives using payment rules.
Overcoming both of these obstacles simultaneously is the subject of \textit{algorithmic mechanism design} \citep{Nisan01algorithmicmechanism}.

In this paper, we propose a solution by which the two obstacles, maximizing the social welfare of GAP as well as extracting true valuations of bidders, are surmounted. The solution provides bidders with incentives to report their valuations truthfully and runs in polynomial time approximating the social welfare with a provable ratio of at least $1-1/e$.

\subsection{Challenges in Algorithmic Mechanism Design}

In algorithmic mechanism design, a mechanism designer wishes to solve an optimization problem, but the inputs to this problem are the private information of self-interested players.
The mechanism designer must thus design a mechanism that solves the optimization problem while encouraging the bidders to truthfully reveal their information.
The game-theoretic concept of truthfulness guarantees that a bidder is better off truthfully interacting with the mechanism regardless of what the other bidders do.

The well-known \textit{Vickrey-Clarke-Groves} (VCG) technique provides truthfulness as well as social welfare maximization in every combinatorial auction. 
The VCG technique, however, is applicable only when the optimal social welfare can be computed to optimality.
Yet, in many cases, including our problem, optimizing social welfare is computationally intractable which makes the VCG technique inapplicable.
Usually, when faced with computational intractability, computer scientists turn to approximations or heuristics.
The VCG technique, unfortunately, cannot be directly applied to approximate solutions \citep{nisan2007computationally}.
In order to resolve the clash between approximation and truthfulness, the maximal-in-distributional-range (MIDR) allocation rules are introduced.

MIDR is the only known general approach for designing randomized truthful mechanisms.
An MIDR algorithm fixes a set of distributions over feasible solutions (the distributional range) independently of the valuations reported by
the self-interested players, and outputs a random sample from the distribution that maximizes expected (reported) welfare \citep{Dobzinski09}.
The best option for a mechanism designer is to devise a MIDR containing an approximation of the optimal social welfare that (very closely) matches the best approximation guarantee known for the problem for which the underlying data are publicly known. 
Finding this type of MIDR or designing an approximation truthful mechanism is not always possible.
Several authors have shown that it is impossible to achieve the same approximation factor in truthful mechanisms ~\citep{Lavi03,papadimitriou2008hardness,dobzinski2013communication,dughmi2015limitations, dobzinski2012computational}.

Looking more closely at the approximation algorithms presented for GAP, we observe that no algorithm can serve as a MIDR allocation rule, although GAP without incentives has been studied extensively in the literature.
\citet{chekuri2005polynomial} explicitly state that the algorithm of 
\citet{shmoys1993approximation} can be adapted to provide a $2$-approximation.
Later, 
\citet{fleischer2006tight} improved the factor to $1-1/e$.
Using a reduction to submodular maximization subject to a matroid constraint, 
\citet{calinescu2011maximizing} 
achieved a ratio of $1-1/e-o(1)$ without using the ellipsoid method which was pivotal in the work done by \citet{fleischer2006tight}. 
An algorithm due to 
\citet{feige2006approximation} yields an approximation factor of $1-1/e+\rho$, $\rho \approx 10^{-180}$ which is the best given approximation ratio for GAP. 
\citet{chakrabarty2010approximability} provide the best-known hardness result showing it is NP-hard to approximate GAP to any factor better than $10/11$.

According to our observations, all foregoing approximation algorithms comprise two algorithms: a \textit{relaxation algorithm} and a \textit{rounding algorithm}.
In order to devise truthful mechanisms, 
\citet{dughmi2011convex} propose an approach which optimizes directly on the outcome of the
rounding algorithm, rather than over the outcome of the relaxation algorithm. 
Since the rounding procedure is embedded into the objective function, this approach is not always computationally tractable.
Assuming the optimization problem can be solved efficiently and the rounding scheme is independent of bidders' valuations, this approach always leads to a MIDR algorithm and is referred to as \textit{convex rounding}.

\citet{lavi2011truthful} propose a framework for deriving MIDR mechanisms from linear programming relaxations. 
They solve the relaxed problem in the first step and then use a special rounding method (convex decomposition) to obtain a randomized integral allocation.
Although Lavi and Swamy also use the common composition of relaxation and rounding algorithms, 
their special rounding procedure produces an expected allocation which is always identical to the scaled down input of the rounding algorithm, component-wise.
Of interest, the rounding procedure used by Lavi and Swamy, guarantees truthfulness-in-expectation. 
Designing truthful mechanisms using the framework of Lavi and Swamy for a given problem is straightforward, however, this type of mechanisms is slow in practice and requires many black-box invocations of an existing approximation algorithm for the problem.
Very recently, \citet{azar2015truthful} present a truthful-in-expectation $1/2$-approximation algorithm for GAP with private values using the framework proposed by Lavi and Swamy and a new rounding technique.


\subsection{Results and Techniques}

It is possible to use the framework developed by \citet{lavi2011truthful} to design a truthful-in-expectation $1/2$-approximation mechanism for GAP; in order to guarantee an improved approximation ratio as well as a higher performance, we follow the convex rounding technique.
The main challenge in using convex rounding is to design an appropriate rounding scheme which induces a convex optimization problem.
Moreover, the rounding scheme should return a feasible solution containing a good approximation of the fractional value.

We design a rounding algorithm with the desired properties for GAP. 
Using the rounding algorithm to obtain a MIDR, we directly optimize over the outcome of the rounding procedure rather than over the outcome of the relaxation algorithm. 
Using this technique, we formulate GAP as a convex optimization problem
where the objective function equals the expected value of a rounding procedure.
In contrast to \citet{dughmi2011convex}, our rounding algorithm uses some information from bidders' valuations.
Our design does not violate the truthfulness since we use  bidders' values solely to search for the optimum in a subset of the range containing the optimal solution, as explained in  Section \ref{sec:truthful}.
This design can be viewed as slightly extending the convex rounding technique and is of independent interest.

We supplement the allocation rule with a payment rule which allows the guarantee of \textit{non-negativity of payments} and \textit{individual rationality} \textit{ex post} rather than providing these important properties only \textit{ex ante}.

The approximation ratio of our mechanism very closely matches the best approximation ratio presented for GAP with publicly known valuations.
In particular, the proposed convex program contains $1-1/e$ ratio of optimum while the best presented approximation ratio of non-truthful algorithms is $1-1/e+\rho$, $\rho \approx 10^{-180}$.

In order to solve the convex program, we present a fractional local search algorithm which approximates the proposed convex optimization problem within an arbitrarily small error, in the sense of an FPTAS.
This leads to an approximate MIDR.

\begin{theorem} \label{thm-eps-midr-gap}
There is a $(1-\epsilon)$-MIDR allocation rule that achieves a $(1-1/e-\epsilon)$-approximation to the social welfare of the generalized assignment problem, for every $\epsilon = 1/{poly}(n)$.
\end{theorem}

\citet{dughmi2011approximately} show how to transform an approximately MIDR allocation rule to an approximately truthful-in-expectation mechanism (see Definition \ref{def-tie}). 
With this black box transformation, we obtain the following.

\begin{theorem} \label{thm-eps-tie-gap}
There is a $(1-\epsilon)$-truthful-in-expectation mechanism that achieves a $(1-1/e-\epsilon)$-approximation to the social welfare of the generalized assignment problem, for every $\epsilon = 1/{poly}(m,n)$.
\end{theorem}

From an algorithmic point of view, the proposed algorithm has advantages over the previously known optimization algorithms for GAP in terms of runtime and simplicity.
We do not employ the ellipsoid method which is identified as pivotal in the work of \citet{fleischer2006tight}.
Our algorithm improves on the one proposed by \citet{calinescu2011maximizing}.
In each iteration of the algorithm by \citet{calinescu2011maximizing}, a random sampling is required to compute the residual increase of assigning an item to a bin which subsequently increases runtime.
The residual increase is treated as an approximate evaluation of gradient of the objective function at a point.  
This residual increase is in fact calculated by taking the average of $(mn)^5$ independent samples, where $m$ and $n$ are the number of items and bins, respectively. 
We use a novel objective function which is specified exactly, rather than by random sampling, whereby it is possible to explicitly calculate the gradient of the objective function which helps to simplify the algorithm and improve the runtime.
It should be noted that in our design, we benefited from the ideas developed in \citet{fleischer2006tight}, \citet{calinescu2011maximizing}, and \citet{dughmi2011approximately}.

\subsection{Paper Structure}
In Section \ref{sec:pre} we introduce necessary notation and definitions used throughout the paper. 
In Section \ref{sec:midr} and Section \ref{sec:payments} we present the MIDR allocation rule and the payment rule, respectively for the setting where bins are held by strategic bidders.
Section \ref{sec:truthful} explains why the presented mechanism is truthful.
The required modification of the mechanism for the case where items are held by bidders is explained in Section \ref{sec:item-bidders}.
Finally, in Section \ref{sec:conclusion} we conclude with a summary and a discussion about future research questions.

\section{Preliminaries} \label{sec:pre}
In the generalized assignment problem, there are $n$ bins, $I$, and $m$ items, $J$. 
Let $v_{ij}$ denote the value of bin $i$ for item $j$.
Each bin $i$ has a different weight $w_{ij}$ for each item $j$ and has a limited capacity $C_i$.
Let $\mathcal{F}_i$ denote the collection of all feasible assignments to bin $i$  $(\forall S\in \mathcal{F}_i: \sum_{j \in S}w_{ij} \leq C_i)$. 
Each item may be assigned to at most one bin.
In the final allocation, some items may remain unassigned.

We assume weights and capacities are publicly known, yet values of assigning items to bins are the private information of bidders.
More formally, we assume values $\{v_{ij}\}_{i \in I, j\in J}$ are private information of bidders.
In the following, we assume \textit{bins} are held by bidders and thus each bidder $i$ has private valuations $\{v_{ij}\}_{j \in J}$. 
We analyze the case where \textit{items} are held by bidders and each bidder $j$ has private valuations $\{v_{ij}\}_{i \in I}$ in Section \ref{sec:item-bidders}.

An allocation $(S_1,\ldots,S_n)$, where $S_i \subseteq J$ denotes the subset assigned to bin $i$, is feasible if 
$\forall i\in I: S_i \in \mathcal{F}_i$ and $\{S_i\}_{i\in I}$ are mutually disjoint. 
The valuation of bin $i$ is defined as $g_i : 2^J \rightarrow \mathbb{R}_{\geq 0}$ 
such that $g_i(S)=\sum_{j \in S} v_{ij}$ if $S\in \mathcal{F}_i$, else $g_i(S)=0$.
With an slight abuse of notation, we sometimes use $g_i(S)$ instead of $g_i(S_i)$, where $S=(S_1,\ldots,S_i,\ldots,S_n).$
The social welfare obtained from a feasible allocation $(S_1,\ldots,S_n)$ is $\sum_{i \in I} g_i(S_i)$. 
The goal is to find a feasible allocation of maximum total social welfare.

In light of the revelation principle, we limit our attention to direct revelation mechanisms. 
Every mechanism has two main components: an \textit{allocation rule} and a \textit{payment rule}.
The allocation rule $\mathcal{A}$ is a function which maps a reported valuation $v=(v_1,\ldots,v_n)$ to an allocation $(S_1,\ldots,S_n)$, where $\forall i: v_i=(v_{ij})_{j\in J}$.
The payment rule is a function from reported valuations to a required payment from each bidder. Let $p_i$ denote the payment rule function for bidder $i$.

\begin{definition}[Maximal in Distributional Range (MIDR)] \label{def-midr}
Given reported valuations $v_1,\ldots, v_n$, and a previously-defined probability distribution over feasible sets $\mathcal{R}$, 
a MIDR returns an outcome sampled randomly from a distribution $D^*\in \mathcal{R}$ 
that maximizes the expected welfare $\E_{x\sim D}[\sum_{i}g_i(x)]$ over all distributions $D \in \mathcal{R}$ \citep{Dobzinski09}.  
\end{definition}
 
Analogously, we define $(1-\epsilon)$-MIDR as follows. 
\begin{definition}[$(1-\epsilon)$-MIDR]
Given  reported valuations $v_1,\ldots, v_n$, and a previously-defined probability distribution over feasible sets $\mathcal{R}$, 
a $(1-\epsilon)$-MIDR returns an outcome sampled randomly from a distribution $D^*\in \mathcal{R}$ 
that $(1-\epsilon)$-approximately maximizes the expected welfare $\E_{x\sim D}[\sum_{i}g_i(x)]$ over all distributions $D \in \mathcal{R}$.  
\end{definition}

An approximately truthful-in-expectation mechanism is defined as follows.
\begin{definition}[$(1-\epsilon)$ truthful-in-expectation] \label{def-tie}
A mechanism is $(1-\epsilon)$-approximately truthful-in-expectation for GAP if, for every bidder $i$, (true) valuation function $v_i$ , (reported) valuation
function $v'_i$, and (reported) valuation functions $v_{-i}$ of the other bidders,
\begin{equation} \label{eq-eps-tie}
\E[g_i(\mathcal{A}(v_i,v_{-i}))-p_i(v_i,v_{-i})] \geq (1-\epsilon)\E[g_i(\mathcal{A}(v'_i,v_{-i}))-p_i(v'_i,v_{-i})].
\end{equation}
The expectation in (\ref{eq-eps-tie}) is taken over the coin flips of the mechanism.
\end{definition}

The goal of our work is to find an allocation and payment rule which constitute a truthful-in-expectation mechanism for GAP and approximates the social welfare as much as possible.

\section{MIDR Allocation Rule for GAP} \label{sec:midr}
We optimize directly over the expected value of the allocation produced by a rounding algorithm.
We let the relaxed feasible set be $\mathcal{R}$ as follows: 
given a vector $x \in \{0,1\}^{I\times 2^J}$, let $x_{i,S}$ indicate whether subset $S$ is assigned to bin $i$.

\begin{equation*}
\displaystyle \mathcal{R}=\bigg\{x \in [0,1]^{I\times 2^J}\hspace{3pt} | 
\forall i:\sum_{S \in \mathcal{F}_i}x_{i,S} \leq 1;
\forall i \in I, \forall S \in \mathcal{F}_i : x_{i,S} \geq 0
\bigg\}.
\end{equation*}

In $\mathcal{R}$ one randomized feasible set is assigned to each bin $i$. 
The sets assigned to different bins may overlap, however in the rounding step each item is assigned only once.
Our intent is to maximize the expected value of the rounded allocation over relaxed set $\mathcal{R}$.
This leads to a MIDR allocation, as explained in Section \ref{sec:truthful}. 
Let us call the rounding algorithm $r_{\text{greedy}}$. 
Algorithm \ref{alg-gap-midr} presents the desired MIDR algorithm.

\begin{algorithm}[H]\label{alg-gap-midr}

\KwData{$v=(v_{ij})_{i\in I, j\in J}$.}
\KwResult{Feasible allocation $(S_1,\ldots,S_n)$.}

	1. Let $x^*$ maximize $\E_{(S_1,\ldots,S_n) \sim r_{\text{greedy}}(x)}[\sum_{i\in I} g_i(S_i)]$ over $x \in \mathcal{R}$. 
	
	2. Let $(S_1,\ldots,S_n) \sim r_{\text{greedy}}(x^*)$.
	
\caption{MIDR allocation rule for the generalized assignment problem.}
\end{algorithm}


Following is a step-by-step procedure to implement Algorithm \ref{alg-gap-midr} and a presentation of the benefits of the outcome of the algorithm.
We start by explaining the rounding algorithm.

\subsection{Greedy Rounding} \label{subsec:rounding}
We choose a rounding algorithm which preserves a good ratio of the fractional solution and returns a feasible allocation in which each item is assigned only once.
We first define helper function $\phi(\cdot)$ which maps a point in $\mathcal{R}$ to a point in $[0,1]^{I \times J}$. 
Let $\phi : \mathcal{R} \rightarrow [0,1]^{I\times J}$ be such that $y=\phi(x)$ iff $\forall i\in I, \forall j\in J:y_{ij}=\sum_{S : j\in S}x_{i,S}$.

The rounding procedure, defined as Algorithm \ref{alg-greedy-round}, has two steps. 
In the first step, given a point $x \in \mathcal{R}$ the rounding procedure finds another point $x' \in \mathcal{R}$ such that
$\forall i\in I, \forall j \in J: y'_{ij}=1-e^{-y_{ij}}$, where $y=\phi(x)$ and $y'=\phi(x')$.
In the second step, the rounding procedure assigns subset $S$ to bin $i$ with probability $x'_{i,S}$ while resolving conflicts as explained in Algorithm \ref{alg-greedy-round}.

We propose Algorithm \ref{alg-packing-property} to perform the first step. 
Algorithm \ref{alg-packing-property} takes a point $x \in \mathcal{R}$ and a desired vector $y' \in [0,1]^{I\times J}$, where $y' \preceq \phi(x)$
and returns another point $x' \in \mathcal{R}$ such that $y'=\phi(x')$.

\begin{algorithm}[H] \label{alg-packing-property}
\SetAlgoNoLine
\SetAlgoNoEnd
\KwData{$x \in \mathcal{R}$, and $y' \in [0,1]^{I\times J}$ such that $y' \preceq \phi(x)$.}
\KwResult{$x' \in \mathcal{R}$ such that $y'=\phi(x')$.}

	Initialize $x' := x$; $\delta = \phi(x')-y'$, where $\delta \in [0,1]^{I\times J}$.
	
	\ForEach {bin $i$} {
		\ForEach {item $j$} {
			
			\Repeat {$\delta_{ij} = 0$}{
				  Choose $x'_{i,S: j \in S}>0$, arbitrarily; 
				  
				  \If {$x'_{i,S} < \delta_{ij}$} {
				  	$\delta_{ij} := \delta_{ij}-x'_{i,S}$; \\
				  	if $S\setminus \{j\} \neq \emptyset$ then $x'_{i, S\setminus \{j\}} := x'_{i, S\setminus \{j\}} + x'_{i,S}$; \\
				  	$x'_{i,S} := 0$;
				  }
				  \Else {
				  	$x'_{i,S} := x'_{i,S} - \delta_{ij}$; \\
				  	if $S\setminus \{j\} \neq \emptyset$ then $x'_{i, S\setminus \{j\}} := x'_{i, S\setminus \{j\}} + \delta_{ij}$; \\
				  	$\delta_{ij} := 0$;
				  }
			}
		}
	}
	\Return {$x'$.}
\caption{An oblivious method for finding a dominated point in $\mathcal{R}$.}
\end{algorithm}

Algorithm \ref{alg-packing-property} returns the desired outcome as confirmed by the following lemma.
\begin{lemma} \label{lem-alg-pack-prop}
Suppose $x \in \mathcal{R}$ with polynomially-many $x_{i,S} > 0$, and $y'  \in [0,1]^{I\times J}$ such that $y' \preceq \phi(x)$. 
If we call Algorithm \ref{alg-packing-property} on $x$ and $y'$, it returns $x' \in \mathcal{R}$ such that $\phi(x')=y'$ with only polynomially-many $x'_{i,S} > 0$. 
\end{lemma}
\proof{}
If the algorithm terminates, we will have $\forall i\in I$, $\forall j \in J$: $\delta_{ij}=0$, and therefore $y'=\phi(x')$.
Thus, we only need to show that the algorithm terminates in polynomial time and $x'$ has polynomially-many positive components.
We show the termination of the algorithm for one bin and one item and since the number of items and bins is polynomial, we obtain the desired conclusion.

Fix bin $i$ and item $j$. 
We consider one iteration in which $x'_{i,S}$ with $j \in S$ is chosen.
Two cases can occur. 
First, $x'_{i,S} < \delta_{ij}$. 
In this case, the number of positive components in $x'$ does not increase, since $x_{i,S}$ becomes zero and at most another positive component is added: $x'_{i,S\setminus \{j\}}$.
This case can occur as many times as the number of $x_{i,S: j \in S}>0$, which are  polynomially-many by assumption.

Second, $x'_{i,S} \geq \delta_{ij}$. In this case, only one new positive component may be added: $x'_{i,S\setminus \{j\}}$.
However, this case can happen only once for item $j$, as $\delta_{ij}$ becomes zero in this step.

Thus, in total for bin $i$ and item $j$, only one new positive component might be included in $x'$ compared to $x$ and the number of iterations is polynomial.
This completes the proof.
\endproof

Thus, for the first step of the rounding algorithm, we call Algorithm \ref{alg-packing-property} on inputs $x$ and $y' \in [0,1]^{I\times J}$ 
where $\forall i \in I, \forall j \in J$: $y'_{ij} = 1-e^{-y_{ij}}$ and $y=\phi(x)$, to obtain the desired point in $\mathcal{R}$.
Notice, that $y' \preceq y$, as needed by Algorithm \ref{alg-packing-property}.

The following is a presentation of the greedy rounding algorithm, $r_{\text{greedy}}$.

\begin{algorithm}[H] \label{alg-greedy-round}

\KwData{$x\in \mathcal{R}$ with polynomially-many $x_{i,S}>0$, $v=(v_{ij})_{i\in I, j\in J}$.}
\KwResult{Feasible allocation $(S_1,\ldots,S_n)$.}

	1. Let $y=\phi(x)$. Let $y' \in [0,1]^{I\times J}$ be such that $y'_{ij} = 1-e^{-y_{ij}}$.
	Invoke Algorithm ($\ref{alg-packing-property}$) with $x$ and $y'$ as the inputs and let $x'$ be the result.

	2. Assign set $S$ to $i$ with probability $x'_{i,S}$ independently for each bin $i$. 
	If some item $j$ is assigned to more than one bin, assign it to the bin among those bins with the maximum value $v_{ij}$.
	Let $S_i$ be the set assigned to bin $i$. 
	
	\Return {$(S_1,\ldots,S_n)$.}
\caption{Greedy rounding algorithm, $r_{\text{greedy}}$.}
\end{algorithm}

In order to analyze the performance of the rounding algorithm, we define a new function.
\begin{equation*}
\begin{array}{ll}
F:[0,1]^{I\times J} \rightarrow \mathbb{R}_{\geq 0} \\
\displaystyle F(y)= \sum_{j=1}^m \sum_{i=1}^n \big( v_{\sigma_j(i),j}-v_{\sigma_j(i+1),j} \big) \big(1-\exp ( {-\sum_{k=1}^i y_{\sigma_j(k),j}} )\big).
\end{array}
\end{equation*}

Where $\sigma_j:I\rightarrow I$ is a permutation on $I$ such that $v_{\sigma_j(i),j}$ is decreasing (non-increasing) when $i$ runs from $1$ to $n$,
and $v_{\sigma_j(n+1),j}=0$. 

Function $F(\cdot)$ is useful in explaining the quality of the rounding algorithm as shown in the following.

\begin{lemma} \label{lem-obj-func-of-rounding}
$ \displaystyle \forall x\in \mathcal{R}: \E_{(S_1,\ldots,S_n) \sim r_{\text{greedy}}(x)}\Big[\sum_{i\in I} g_i(S_i)\Big] = F(\phi(x))$.
\end{lemma}
\proof{}
Assume $x \in \mathcal{R}$. 
Let $x'$ be the outcome of Step 1 of Algorithm \ref{alg-greedy-round}. Let $y=\phi(x)$ and $y'=\phi(x')$.
We calculate the expected value achieved from the assignment of item $j$ in the integral allocation.

Fix item $j$. For simplicity, we assume that $\sigma_j(i)=i$. 
This means that bins with smaller indices have higher valuations for $j$.
We find the expected value returned from item $j$;  for other items, the argument is similar.
With probability $y'_{1j}$ the set assigned to bin $1$ contains $j$ thus $j$ is assigned to $1$.
Recall that $y'_{1j}=\sum_{S:j\in S} x'_{1,S}$.
Therefore, with probability $y'_{1j}$, the value of returned allocation is $v_{1j}$.
With probability $(1-y'_{1j})y'_{2j}$ the set assigned to bin $1$ does not contain the item but the set assigned to bin $2$ contains the item and therefore item $j$ is assigned to bin $2$. 
This case leads to a returned value of $(1-y'_{1j})y'_{2j} v_{2j}$.

Continuing in a similar manner for other bins, the achievable expected value becomes
$y'_{1j} v_{1j}+(1-y'_{1j})y'_{2j} v_{2j} + \ldots + \prod_{k=1}^{n-1}(1-y'_{kj}) y'_{nj} v_{nj}$, which in turn
equals $\sum_{i=1}^n (v_{ij}-v_{i+1,j}) (1-\prod_{k=1}^{i}(1-y'_{kj}))$. 
The equality of the two terms can be observed by simply extending the latter.
Taking into account that $y'_{ij} = 1-e^{-y_{ij}}$, by summing over all items we obtain the desired conclusion, using linearity of expectation.
\endproof

Therefore, we need to optimize $F(\phi(x))$ over $x \in \mathcal{R}$. 
Optimizing $F(\phi(x))$ over $x \in \mathcal{R}$ is essentially the same as optimizing $F(y)$ over $y \in \mathcal{P}$, 
where 
\begin{equation*}
\mathcal{P}=\bigg\{y \in [0,1]^{I\times J}\hspace{3pt} |\hspace{3pt} y=\phi(x) \hspace{3pt}\&\hspace{3pt} x \in \mathcal{R} \bigg\}.
\end{equation*}
As a result, what remains is to explain how to solve $\max_{y \in \mathcal{P}}F(y)$, and the quality of the solution.

\subsection{The Approximation Ratio}
We show the quality of our method by comparing $\max_{y \in \mathcal{P}}F(y)$ to the optimal solution to the configuration LP of GAP.
The configuration LP of GAP is as follows:
  \begin{alignat*}{2}
    \text{GAP-CLP:} & \\
    \text{max}   & \sum_{i\in I, S \in \mathcal{F}_i} x_{i,S} g_i(S)  \\
	 & \forall j \in J : \sum_{i \in I, S \in \mathcal{F}_i : j \in S} x_{i,S} \leq 1, \ & \\
	 &\forall i \in I : \sum_{S \in \mathcal{F}_i} x_{i,S} \leq 1, \ &   \\
	 & \forall i \in I, \forall S \in \mathcal{F}_i : x_{i,S} \ge 0, \ &  
  \end{alignat*}

To be able to compare GAP-CLP to $F(y)$, we first introduce a new variable into the program and then rearrange the objective function.
Let $y \in [0,1]^{I\times J}$ be such that $\forall i \in I, \forall j\in J:y_{ij}=\sum_{S \in \mathcal{F}_i : j\in S}x_{i,S}$.
Using this new variable we define polytope $\mathcal{P}'$ as in the following:

\begin{equation*}
\begin{array}{ll}
\displaystyle \mathcal{P}'= &\bigg\{y \in [0,1]^{I\times J}\hspace{3pt} | \\
& \forall j\in J:\sum_{i\in I}y_{ij} \leq 1;   \text{\textbf{(1)}}\\
& \forall i\in I, \forall j\in J:y_{ij}=\sum_{S \in \mathcal{F}_i : j\in S}x_{i,S}; \\
& \forall i:\sum_{S \in \mathcal{F}_i}x_{i,S} \leq 1; \\
& \forall i \in I, \forall S \in \mathcal{F}_i : x_{i,S} \geq 0
\bigg\}.
\end{array}
\end{equation*}

We notice that $\mathcal{P}' \subseteq \mathcal{P}$ since $\mathcal{P}'$ has an additional constraint (Constraint 1).
We rearrange the objective function of GAP-CLP to be a function of items ($y$) rather than subsets, ($x$).

\begin{equation*}
\begin{array}{ll}
\displaystyle \sum_{i \in I, S \in \mathcal{F}_i} x_{i,S} g_i(S) & = 
\displaystyle \sum_{i \in I, S \in \mathcal{F}_i} x_{i,S} \sum_{j\in S} v_{ij} \\
&= \displaystyle \sum_{i \in I, j \in J} v_{ij}\sum_{S \in \mathcal{F}_i : j\in S}x_{i,S} \\
&= \displaystyle  \sum_{i \in I, j \in J} v_{ij}y_{ij} \\

\end{array}
\end{equation*}


Consequently, solving GAP-CLP is equivalent to finding $\max_{y \in \mathcal{P}'} \sum_{i \in I, j \in J} v_{ij}y_{ij}$. 
We are now ready to compare $\max_{y\in \mathcal{P}}F(y)$ with the optimal integral solution to the GAP (denoted by $OPT$).

\begin{lemma} \label{lem-0.63}
$\max_{y\in \mathcal{P}} F(y) \geq (1-\frac{1}{e}) OPT$.
\end{lemma}
\proof{} 
We observe that 

\begin{equation*}
\max_{y\in \mathcal{P}}F(y) \geq \max_{y\in \mathcal{P}'}F(y) \geq (1-\frac{1}{e})\max_{y\in \mathcal{P}'} \sum_{i \in I, j \in J} v_{ij}y_{ij} \geq (1-\frac{1}{e}) OPT.
\end{equation*}

The first inequality holds since $\mathcal{P}' \subseteq \mathcal{P}$. 
The last inequality holds because $\displaystyle \max_{y\in \mathcal{P}'} \sum_{i \in I, j \in J} v_{ij}y_{ij}$ in fact provides a solution to GAP-CLP which is obviously greater than $OPT$.
For the second inequality, consider item $j$ and $y \in \mathcal{P}'$. 
For simplicity, we assume $\forall i: \sigma_j(i) = i$.
We have $\sum_{i=1}^n y_{ij}\leq 1$, since $ y\in \mathcal{P}'$.
Considering the fact that $1-e^{-x} \geq (1-\frac{1}{e})x$ for $x \in [0,1]$, we obtain
\begin{displaymath}
\begin{array}{lll}
(v_{1j}-v_{2j})(1-e^{-y_{1j}}) & \geq (v_{1j}-v_{2j})(1-\frac{1}{e}) y_{1j} \\
(v_{2j}-v_{3j})(1-e^{-y_{1j}-y_{2j}}) & \geq (v_{2j}-v_{3j}) (1-\frac{1}{e}) (y_{1j}+y_{2j}) \\
\ldots \\
(v_{n-1,j}-v_{nj})(1-e^{-\sum_{k=1}^{n-1}y_{kj}}) & \geq (v_{n-1,j}-v_{nj}) (1-\frac{1}{e}) (\sum_{k=1}^{n-1}y_{kj}) \\
(v_{nj})(1-e^{-\sum_{k=1}^{n}y_{kj}}) & \geq (v_{nj}) (1-\frac{1}{e}) (\sum_{k=1}^{n}y_{kj}) \\
\end{array}
\end{displaymath}
Summing both sides, we obtain

\begin{equation*}
\sum_{i =1}^n \big(v_{i,j}-v_{i+1,j}\big) \big(1-\exp({-\sum_{k=1}^i y_{k,j}})\big) \geq 
\big(1-\frac{1}{e}\big){\sum_{i \in I} v_{ij} y_{ij}}.
\end{equation*}

Obtaining this inequality for all items then, and summing them up, we obtain the desired conclusion.
\endproof

Thus, what remains is to show how to maximize $F(y)$ over $y\in \mathcal{P}$, the topic of Section \ref{sec:solve-cx-problem}.

\subsection{Solving the Convex Optimization Problem} \label{sec:solve-cx-problem}
We wish to solve $\max_{y \in \mathcal{P}} F(y)$ which is essentially equivalent to the following mathematical optimization problem: 

  \begin{alignat*}{2}
    \text{GAP-CP:} & \\
    \text{Maximize}   & \sum_{j=1}^m \sum_{i=1}^n \big(v_{\sigma_j(i),j}-v_{\sigma_j(i+1),j}\big) \big(1-\exp({-\sum_{k=1}^i y_{\sigma_j(k),j}})\big)  \\
	 &\forall i\in I, \forall j \in J : \sum_{S \in \mathcal{F}_i : j \in S} x_{i,S} = y_{ij},\\
	 &\forall i \in I : \sum_{S \in \mathcal{F}_i} x_{i,S} \leq 1, \ &   \\
	 & \forall i \in I, \forall S \in \mathcal{F}_i : x_{i,S} \ge 0. \ &  
  \end{alignat*}

First, we show that GAP-CP is a convex optimization problem. 
All constraints in the program are linear thus we only need to show that the objective function, $F(y)$, is concave/convex as shown by the following theorem.
\begin{lemma} \label{lem-concave-func}
$F(y)$ is a concave function.
\end{lemma}
\proof{}
$F(y)$  is concave in $y$, because it is a non-negative weighted sum of functions which are compositions of the concave function $1-e^{-x}$ with affine function $x \rightarrow \sum_{k=1}^i y_{\sigma_j(k),j}$ \cite{boyd2009convex}. 
\endproof

In order to solve the convex optimization problem, we present a fractional local search algorithm.
Our algorithm gets arbitrarily close to the optimal solution.
The difficulty in solving the convex optimization problem mostly arises from the exponential number of variables in the convex program. 
As a result, we are able to implement a $(1-\epsilon)$-MIDR allocation rule, for any $\epsilon = 1/\text{poly}(n)$.

Our algorithm employs a polynomial number of iterations to get as close as a predefined precision to the optimal solution. 
In every iteration of the algorithm, we need to find $y^* \in \mathcal{P}$ which maximizes $y\cdot \nabla F(y)$ 
\footnote{We remind the reader that $\nabla F$, the gradient of F, is a vector whose coordinates are the first partial derivatives $\frac{\partial{F}}{\partial y_{ij}}$. We denote by $\frac{\partial{F}}{\partial y_{ij}}\big |_{y}$ the gradient coordinate $(i,j)$ evaluated at point $y$.} 
over all $y \in \mathcal{P}$.
According to Proposition \ref{prop-linear-max}, maximizing $y \cdot v$ over all $y \in \mathcal{P}$ for every cost function $v$, is equivalent to finding set $S^*_i \in \mathcal{F}_i$ for every bin $i$ which maximizes $\sum_{j\in S^*_i} v_{ij}$.

\begin{proposition} \label{prop-linear-max}
$\max_{y \in \mathcal{P}} \sum_{i \in I, j \in J}  v_{ij}y_{ij} = \sum_{i\in I} \max\{\sum_{j \in S} v_{ij} : S \in \mathcal{F}_i\}$.
\end{proposition}
\proof
\begin{displaymath}
\begin{array}{ll}
\displaystyle \max_{y \in \mathcal{P}} \sum_{i \in I, j \in J} v_{ij}y_{ij} &= 
\displaystyle \max_{x \in \mathcal{R}} \sum_{i\in I, j\in J} \Big(v_{ij}\sum_{S \in \mathcal{F}_i : j\in S}x_{i,S}\Big)\\ 
& =\displaystyle \max_{x \in \mathcal{R}} \sum_{i \in I} \sum_{S \in \mathcal{F}_i} \Big(x_{i,S} \sum_{j \in S} v_{ij} \Big)\\
& = \displaystyle \sum_{i \in I} \max\{\sum_{j \in S} v_{ij} : S \in \mathcal{F}_i\}.
\end{array}
\end{displaymath}

The first equality holds since for every $y \in \mathcal{P}$, there exists $x \in \mathcal{R}$ where $y=\phi(x)$.
The last equality holds since if $x \in \mathcal{R}$ then $\sum_{S \in \mathcal{F}_i} x_{i,S} \leq 1$.  
\endproof

Finding $\max\{\sum_{j \in S} v_{ij} : S \in \mathcal{F}_i\}$ is essentially solving a knapsack subproblem for bin $i$. 
To do so, we invoke the FPTAS for the knapsack problem. 
We say for any $v_i=(v_{ij})_{j \in J}$ and $0<\epsilon<1$, \emph{\textbf{KnapsackFptas}}$(v_i, \epsilon)$ returns subset $S_i \in \mathcal{F}_i$ where $\sum_{j\in S_i}v_{ij} > (1-\epsilon) \max\{\sum_{j\in S} v_{ij} : S\in \mathcal{F}_i\}$.

We store the computed vector in each iteration in a set $\mathcal{Z}$. 
We keep the size of $\mathcal{Z}$ to be of at most $\frac{1}{\delta}$ for $\delta = \frac{1}{n}$; $\delta$ will be defined later.
As long as $|\mathcal{Z}| < \frac{1}{\delta}$ we simply add the current vector to $\mathcal{Z}$.
When $|\mathcal{Z}| = \frac{1}{\delta}$, in each iteration we add the current vector and remove one vector from $\mathcal{Z}$ which has the least value with respect to the current gradient.
The solution returned by the algorithm, $x$, has the property that $y=\phi(x)$ is a convex combination of the vectors in $\mathcal{Z}$: $y=\delta \cdot \sum_{z\in \mathcal{Z}} z$.
We continue updating $\mathcal{Z}$ until the increase in $F(y)$ is below a predefined threshold.

Now, we present the main algorithm.
Let $M$ denote $\max\{v_{ij} : i\in I; j \in J\}$.
  
\begin{algorithm}[H]\label{alg-frac-local-search}
\SetAlgoNoEnd

\KwData{$v=(v_{ij})_{i\in I, j\in J}$, $0 < \epsilon \leq 1/n$.}
\KwResult{$x \in \mathcal{R}$ such that $F(\phi(x)) \geq (1-o(1)) \max \{F(y)|y \in \mathcal{P}\}$.}

	0. Initialize $x:=\vec{0}$; $y:=\vec{0}$;  
	$\mathcal{Z} = \emptyset$;
	$\delta=\frac{\epsilon}{6mn^2}$;
	
		1. $u := \nabla F(y)$; 	$z:= \vec{0}$; \tcc*{\footnotesize $x \in [0,1]^{I\times 2^J}$ and $y,u,z\in [0,1]^{I\times J}$} 

		2. \ForEach {bin $i$} {
		Let $S := $\emph{\textbf{KnapsackFptas}}$(u_i, \epsilon)$;
		for all $j \in S$ update $z_{ij} := 1$; 
		}
		
		3. \If {$ (z - y) \cdot \nabla F(y) > \epsilon M $} {
		
			\If {$(|\mathcal{Z}| < \frac{1}{\delta})$} {			
					
				Update $y := y+\delta z$; $\mathcal{Z}:=\mathcal{Z}\cup \{z\}$; 
		
			}
			\Else {
				Update $y := y+\delta (z-z^{min})$, where $\displaystyle z^{min} = \argmin_{z'\in \mathcal{Z}} z' \cdot u$; 
				$\mathcal{Z} := (\mathcal{Z}\setminus \{z^{min}\}) \cup \{z\}$;
		
			}
			Go back to Step 1;
		} 
		4. \ForEach {$z \in \mathcal{Z}$} {
			\ForEach {bin $i$} {
				Update $x_{i,S} := x_{i,S}+\delta$, where $S=\{j|z_{ij}=1\}$;
			}
		}
	\Return {$x$.}
\caption{Fractional local search algorithm}
\end{algorithm}

\begin{lemma} \label{lem-x-in-polytope}
Algorithm \ref{alg-frac-local-search} produces a solution $x$ such that $x \in \mathcal{R}$.
\end{lemma}
\proof{}
We observe that the set $\mathcal{Z}$ contains at most $\frac{1}{\delta}$ elements;
as long as $|\mathcal{Z}| < \frac{1}{\delta}$, one element $z$ is included into the set and when $|\mathcal{Z}| = \frac{1}{\delta}$, one element is added and one element is removed from the set.

Towards the end of Algorithm \ref{alg-frac-local-search} (Step 4), for each $z \in \mathcal{Z}$ and each bin $i$, one positive component ($x_{i,S}$) is increased up to $\delta$ which in turn means for each bin $i$ we have $\sum_{S \in \mathcal{F}_i} x_{i,S} \leq \frac{1}{\delta} \cdot \delta = 1$.
That means $x \in \mathcal{R}$, the desired conclusion.
\endproof

\begin{lemma} \label{lem-frac-alg-rario}
Algorithm \ref{alg-frac-local-search} returns $x\in \mathcal{R}$ such that $F(\phi(x)) \geq (1-o(1)) \max \{F(y)|y \in \mathcal{P}\}$.
\end{lemma}
\proof{}
Assume $x$ is the outcome of Algorithm \ref{alg-frac-local-search}. 
According to Lemma \ref{lem-x-in-polytope}, $x \in \mathcal{R}$.
Let $y=\phi(x)$.
Let $z$ be the calculated vector in the last iteration in Step 2, i.e. $(z-y) \cdot \nabla F(y) \leq \epsilon M$. 

Let $y^* = \arg \max_{y \in \mathcal{P}}F(y)$.
According to Proposition \ref{prop-linear-max}, $z \cdot \nabla F(y) \geq (1-\epsilon)\max_{w \in \mathcal{P}} w\cdot \nabla F(y)$. 
Hence, $z \cdot \nabla F(y) \geq (1-\epsilon)y^* \cdot \nabla F(y)$. Thus, we get

\begin{equation*}
\begin{array}{ll}
F(y^*)-F(y) & \leq (y^*-y) \cdot \nabla F(y)  \\ 
& \leq \frac{1}{1-\epsilon}(z\cdot \nabla F(y)-y \cdot \nabla F(y))+ \frac{\epsilon}{1-\epsilon} y \cdot \nabla F(y)  \\
& \leq \frac{\epsilon}{1-\epsilon}M+\epsilon F(y^*)  \end{array}
\end{equation*}

The first inequality is because of the concavity of $F$. 
The second inequality is by rearranging and using inequality $z \cdot \nabla F(y) \geq (1-\epsilon)y^* \cdot \nabla F(y)$.
The third inequality holds since $(z-y)\cdot \nabla F(y) \leq \epsilon M$, and $y \cdot \nabla F(y) < (1-\epsilon) F(y^*)$. 
If $y \cdot \nabla F(y) \geq (1-\epsilon) F(y^*)$ then, by concavity of $F$, $F(y) \geq y \cdot \nabla F(y) \geq (1-\epsilon) F(y^*)$, and therefore Lemma \ref{lem-frac-alg-rario} holds.

Now, using $\epsilon = \frac{1}{n}$, we obtain $F(y^*)-F(y) \leq \frac{1}{n-1} M + \frac{1}{n} F(y^*) \leq \frac{2}{n-1} F(y^*)$.
Hence, when Algorithm  \ref{alg-frac-local-search} terminates $F(y) \geq (1-o(1)) F(y^*)$, the desired conclusion.
\endproof

The change in $y$ in each iteration is either $\delta z$ or $\delta (z-z^{min})$ for $|\mathcal{Z}| < \frac{1}{\delta}$ and $|\mathcal{Z}| = \frac{1}{\delta}$, respectively.
The change in gradient, however, has a certain upper bound when $y$ changes by a certain amount, as Lemma \ref{lem-up-low-change-gradient} shows.

\begin{lemma} \label{lem-up-low-change-gradient}
For any $y$ and $y'$ with $||y-y'||_{\infty} \leq \delta$ and any  $i$ and $j$, 
\begin{displaymath}
e^{-n\delta} \cdot \frac{\partial{F}}{\partial y_{ij}}\bigg |_{y} \leq \frac{\partial{F}}{\partial y_{ij}}\bigg |_{y'} \leq e^{n\delta} \cdot \frac{\partial{F}}{\partial y_{ij}}\bigg |_{y}.
\end{displaymath}
\end{lemma}
\proof{}
Consider the gradient of $F$.
For simplicity, we assume that $\sigma_j(i)=i$. 

\begin{equation} \label{gradient-F}
\frac{\partial{F}}{\partial y_{ij}}=\sum_{l=i}^n(v_{lj}-v_{l+1,j}) \exp({-\sum_{k=1}^l y_{kj}}).
\end{equation}

Considering $\sum_{k=1}^l y'_{kj} \leq \sum_{k=1}^n y_{kj}+n\delta$ and
from (\ref{gradient-F}) we obtain $\frac{\partial{F}}{\partial y_{ij}}\bigg |_{y'} \geq 
e^{-n\delta} \cdot \frac{\partial{F}}{\partial y_{ij}}\bigg |_{y}$.

Similarly, from $\sum_{k=1}^l y'_{kj} \geq \sum_{k=1}^n y_{kj}-n\delta$ and (\ref{gradient-F}), we arrive at $\frac{\partial{F}}{\partial y_{ij}}\bigg |_{y'} \leq 
e^{n\delta} \cdot \frac{\partial{F}}{\partial y_{ij}}\bigg |_{y}$. 
This completes the proof.
\endproof

The following lemma is useful in showing the progress of the algorithm.
\begin{lemma} \label{lem-product-neighbor-gradient}
For any $y$ and $y'$ with $||y-y'||_{\infty} \leq \delta$,

\begin{displaymath}
(y'-y)\cdot \nabla F(y') \geq (y'-y)\cdot \nabla F(y)- 3\delta n^2mM.
\end{displaymath}
\end{lemma}

\proof{}
Let $y'-y=z^+-z^-$ where for all $i$ and $j$, $0\leq z_{ij}^+\leq \delta$ and $0\leq z_{ij}^-\leq \delta$.
From Lemma \ref{lem-up-low-change-gradient}, $z^+\cdot \nabla F(y') \geq e^{-n\delta} z^+ \cdot \nabla F(y)$ and $z^- \cdot \nabla F(y') \leq e^{n\delta}z^-\cdot \nabla F(y)$.
From these inequalities and using inequalities $e^{-x} \geq 1-x$ and $e^x \leq 1+2x$ for $0\leq x\leq 1$ and $\delta < 1/n$, we get

\begin{displaymath}
\begin{array}{ll}
(y'-y)\cdot \nabla F(y') &= (z^+-z^-) \cdot \nabla F(y') \\
&\geq (e^{-n\delta}z^+-e^{n\delta} z^-) \cdot \nabla F(y) \\
&\geq ((1-n\delta)z^+-(1+2n\delta)z^-)\cdot \nabla F(y) \\
&=(z^+-z^-)\cdot \nabla F(y)-n\delta (z^++2z^-)\cdot \nabla F(y) \\
&\geq (y'-y)\cdot \nabla F(y)-3\delta n^2mM.
\end{array}
\end{displaymath}

The last inequality holds because for every $z \in [0,1]^{I\times J}$, $z\cdot \nabla F(y) \leq nmM$.
This is true since for any $y$, we have $\frac{\partial{F}}{\partial y_{ij}} \leq M$ and in the best possible case for $z$, every bin packs the $m$ items and produces a value of $mM$.
This completes the proof.
\endproof

\begin{lemma} \label{lem-frac-loc-growth}
In each iteration, the value of $F(y)$ increases by at least $\frac{\epsilon^2}{12mn^2}M$.
\end{lemma}
\proof{}
As long as the algorithm continues we have $(z - y) \cdot \nabla F(y) > \epsilon M$.
First, we consider the case where $|\mathcal{Z}|<\frac{1}{\delta}$. We have 

\begin{displaymath}
\begin{array}{ll}
F(y + \delta z) & \geq F(y) + \delta z \cdot \nabla F(y+\delta z) \\
& \geq F(y) + \delta e^{-n\delta} z \cdot \nabla F(y)\\
& \geq F(y) + \delta e^{-n\delta} \epsilon M \\
\end{array}
\end{displaymath}

The first inequality is because of the concavity of $F$.
The second inequality holds because of Lemma \ref{lem-up-low-change-gradient}.
The third inequality is because $(z - y) \cdot \nabla F(y) > \epsilon M$ implies that $z \cdot \nabla F(y)>\epsilon M$, as we always have $\nabla F(y) \geq \vec{0}$.

Now, using $\delta = \frac{\epsilon}{6mn^2}$, we obtain

\begin{displaymath}
F(y + \delta z) \geq F(y) + \frac{\epsilon^2}{12mn^2}M.
\end{displaymath}

Second, we consider the case where $|\mathcal{Z}|=\frac{1}{\delta}$. We have

\begin{displaymath}
\begin{array}{ll}
F(y + \delta(z-z^{min})) & \geq F(y) + \delta (z-z^{min}) \cdot \nabla F(y+\delta(z-z^{min})) \\
& \geq F(y) + \delta (z-z^{min}) \cdot \nabla F(y) - 3\delta ^2n^2mM\\
& \geq F(y) + \delta\epsilon M- 3\delta ^2mn^2M \\
\end{array}
\end{displaymath}

The first inequality is because of the concavity of $F$.
The second inequality holds because of Lemma \ref{lem-product-neighbor-gradient}.
The third inequality is because $(z-z^{min}) \cdot \nabla F(y) \geq (z - y) \cdot \nabla F(y)$, as shown in the following.

By definition of $z^{min}$, $z^{min} \cdot \nabla F(y) \leq z'\cdot \nabla F(y)$ for all $z' \in \mathcal{Z}$. 
Thus, $|\mathcal{Z}| \cdot z^{min}\cdot \nabla F(y) \leq \sum_{z'\in \mathcal{Z}} z'\cdot \nabla F(y)$,
 which in turn means  $z^{min} \cdot \nabla F(y) \leq y \cdot \nabla F(y)$.
 Observe that $y = \delta \cdot \sum_{z' \in \mathcal{Z}} z'$.

Now, using $\delta = \frac{\epsilon}{6mn^2}$, we obtain

\begin{displaymath}
F(y + \delta(z-z^{min})) \geq F(y) + \frac{\epsilon^2}{12mn^2}M.
\end{displaymath}

This completes the proof.
\endproof

\begin{lemma} \label{lem-max-no-itr}
After at most ${12m^2n^2}/{\epsilon^2}$ iterations, Algorithm \ref{alg-frac-local-search} terminates.
\end{lemma}
\proof{}
Since $M$ denotes $\max\{v_{ij} : i\in I; j \in J\}$, $mM$ is an upper bound for $\max_{y\in \mathcal{P}}F(y)$. 
Recall that  

\begin{equation*}
\max_{y \in \mathcal{P}} F(y) = \max_{x \in \mathcal{R}} \E_{(S_1,\ldots,S_n) \sim r_{\text{greedy}}(x)}\Big[\sum_{i\in I} g_i(S_i)\Big] \leq mM.
\end{equation*}

Based on Lemma \ref{lem-frac-loc-growth}, in each iteration the growth in value is at least $\frac{\epsilon^2}{12mn^2}M$, 
Algorithm \ref{alg-frac-local-search} thus in at most ${12m^2n^2}/{\epsilon^2}$ iterations, reaches the value of $mM$, which is an upper bound on the best solution.
This concludes the proof.
\endproof

We thus achieve a $(1-\epsilon)$-MIDR allocation rule that runs in polynomial time. This concludes the proof of Theorem \ref{thm-eps-midr-gap}.

\subsection{Simplifying the Rounding Procedure}
As we note, it is possible to simplify the rounding procedure (Algorithm \ref{alg-greedy-round}), further. The simplified rounding is as follows.

Given $x\in \mathcal{R}$, let $y=\phi(x)$. We assign set $S$ to each bin $i$ independently with probability $x_{i,S}$.
Next, for each item $j$ we do as follows.
If item $j$ is assigned to bin $i$, we let the bin hold the item with probability $\frac{1-e^{-y_{ij}}}{y_{ij}}$.
This means, we withdraw the item from the bin with the complementary probability $1-\frac{1-e^{-y_{ij}}}{y_{ij}}$ to make sure that the probability of assigning item $j$ to bin $i$ is not $y_{ij}$ but $1-e^{-y_{ij}}$ which is necessary to maintain the MIDR property.
According to the MIDR principle, the expected value of the randomized integral assignment should equal the calculated fractional value.
Finally, if some item $j$ is assigned to more than one bin, we assign it to the bin among those bins with the maximum value $v_{ij}$.

In order to use the allocation rule algorithm (Algorithm \ref{alg-gap-midr}) as an optimization algorithm, one can employ a more simple rounding algorithm.
The simpler rounding requires only Step 2 of Algorithm \ref{alg-greedy-round}.
For an optimization purpose, there is no need to also execute Step 1 of Algorithm \ref{alg-greedy-round}.
Thus, after finding a fractional solution $x$ by invoking Algorithm \ref{alg-frac-local-search}, we assign set $S$ to each bin $i$ with probability $x_{i,S}$ and resolve conflicts according to the technique explained in Algorithm \ref{alg-greedy-round}.
This improves the runtime for the optimization purpose.

\section{Computing Payments} \label{sec:payments}
Supplementing the MIDR allocation rule of Section \ref{sec:midr} with VCG payments yields a truthful-in-expectation mechanism.
We compute payments in order to also enforce non-negativity of payments and individual rationality, \textit{ex post}.

To compute the VCG fractional payment $p^{\textit{frac}}_i$ for bidder $i$, we need to compute two components: first, the Clarke pivot, $h_i(v_{-i})$, which is the best achievable social welfare by bidders other than $i$,  
 and second, the value gained by bidders other than bidder $i$ in the current fractional solution.
We can calculate $h_i(v_{-i})$ by rewriting GAP-CP for the market without bidder $i$, i.e. $v_{ij}=0$ for all $j$.
To compute the value gained by other bidders in the fractional allocation, $F_{-i}(y^*)$, we set $\forall j \in J: v_{ij} = 0$ in $F(y^*)$, assuming that $y^*$ is the outcome of Algorithm \ref{alg-frac-local-search}. 
Function $F(y)$ is explicitly known to us and we can set in it $v_{ij}$ to $0$.
Finally, $p^{\textit{frac}}_i = h_i(v_{-i}) - F_{-i}(y^*)$.

\begin{example} \label{ex-payment}
Consider a setting in which two bidders (1 and 2) have valuations for two items as follows:
$v_{11} = 8, v_{12}=5$ and $v_{21}=4, v_{22}=10$. In this case, 
\begin{equation*}
F(y)=(8-4)(1-e^{-y_{11}})+4(1-e^{-y_{11}-y_{21}})
+(10-5)(1-e^{-y_{22}}) + 5(1-e^{-y_{12}-y_{22}}).
\end{equation*} 

Now, assume $y^*_1=(0.6,0.3)$ and $y^*_2=(0.4,0.7)$. Then,
\begin{equation*}
F_{-1}(y^*)=(0-4)(1-e^{-0.6})+4(1-e^{-1})
+(10-0)(1-e^{-0.7}) + 0(1-e^{-1}).
\end{equation*} 

\end{example}

The value gained by bidder $i$ in the fractional allocation is therefore $w^{\textit{frac}}_i=F(y^*) - F_{-i}(y^*)$.
Assuming that $S_i$ is the subset assigned to bidder $i$ by the rounding procedure, 
we can compute the randomized payment for bidder $i$, $p_i$, satisfying individual rationality and non-negativity of payments as follows.

\begin{displaymath}
  \begin{array}{l l}
 p_i = \left\{ 
  \begin{array}{l l}
    \frac{g_i(S_i)}{w^{\textit{frac}}_i} p^{\textit{frac}}_i& \quad \text{if $w^{\textit{frac}}_i > 0$,}\\
    0 & \quad \text{if $w^{\textit{frac}}_i = 0$}.
  \end{array} \right. &
  \end{array}
\end{displaymath}

\section{Truthfulness} \label{sec:truthful}

It has been proven that if the allocation algorithm of a mechanism is maximal-in-range and the payment rule calculates the payments by applying the same payment idea as in VCG, then the mechanism is truthful \citep{nisan2007computationally}.
Mechanisms using the idea of VCG to calculate payments are termed VCG-based mechanisms.

Obviously, our mechanism is a VCG-based mechanism as we calculate the payments following the idea of VCG. 
In order to show truthfulness, it suffices to show that the allocation algorithm is maximal-in-range.
Since our mechanism is randomized, we need to show that the allocation rule is maximal in distributional range.
To do so, we describe the \textit{distributional range} of allocations over which the allocation rule optimizes social welfare to optimality.
By MIDR definition, the range of allocations has to be chosen before any valuation has been seen.

Let $\Pi$ denote the set of all permutations on all bins.
The rounding algorithm (Algorithm \ref{alg-greedy-round}) actually works with a specific permutation on bins for each item.
In particular, the rounding algorithm uses $\sigma_j$ for each item $j$ which reorders the bins in decreasing order of their values for item $j$.
We recall that $\sigma_j$ is formally defined in Subsection \ref{subsec:rounding}.

We can look at the rounding algorithm as a function which takes the permutation $\pi_j$ for each item $j$ as input.
Let us call this parameterized rounding algorithm $r$ and define it as Algorithm \ref{alg-round}.

\begin{algorithm}[H] \label{alg-round}

\KwData{$x\in \mathcal{R}$, $\pi_j \in \Pi$ for each item $j$.}
\KwResult{Feasible allocation $(S_1,\ldots,S_n)$.}

	1. Let $y=\phi(x)$. Let $y' \in [0,1]^{I\times J}$ be such that $y'_{ij} = 1-e^{-y_{ij}}$.
	Invoke Algorithm ($\ref{alg-packing-property}$) with $x$ and $y'$ as the inputs and let $x'$ be the result.

	2. Independently for each bin $i$, assign set $S$ to $i$ with probability $x'_{i,S}$. 
	If some item $j$ is assigned to more than one bin, then \textit{assign item $j$ to the bin among those bins that precedes others in $\pi_j$}. 
	
	\Return {$(S_1,\ldots,S_n)$.}
\caption{Rounding algorithm, $r$.}
\end{algorithm}

Algorithm \ref{alg-greedy-round} thus can be rewritten as $r_{\text{greedy}}(x)=r(x,\sigma_1,\sigma_2,\ldots,\sigma_m)$.

For each point $x\in \mathcal{R}$,  $r(x,\pi_1,\ldots,\pi_m)$ rounds point $x$ to an integer point by taking into account permutation $\pi_j \in \Pi$ for each item $j$.
We let the \textit{domain} of function $r$ comprise all $x\in \mathcal{R}$ as well as all $\pi_j \in \Pi$, $\forall j\in J$.
The \textit{range} of function $r$ then is the range over which our allocation algorithm optimizes the social welfare.
Formally, we define the range as follows.
\begin{displaymath}
\textit{Range} \equiv \bigcup_{x \in \mathcal{R}, \pi \in \Pi^J} \{r(x,\pi)\}.
\end{displaymath}

The range is clearly independent of private values as it only takes into account points $x$ and the permutations.
Maximizing over this range defines a maximal in distributional range algorithm. 
In order to maximize over the range we don't need to search the full range. 
Rather, it suffices to look for the maximum only in the part of the range containing the maximum.

We utilize this fact in our MIDR.
In particular, in Algorithm \ref{alg-gap-midr} we maximize $r(x,\sigma_1, \ldots, \sigma_m)$ over $x \in \mathcal{R}$.
There is no need to take into account other permutations since the value of $r(x,\sigma_1, \ldots, \sigma_m)$ is always as high as that of $r(x,\pi_1, \ldots, \pi_m)$ where at least for one $j$ $\pi_j \neq \sigma_j$.
To rephrase, maximizing  $r(x,\sigma_1, \ldots, \sigma_m)$ over $x \in \mathcal{R}$ is equivalent to maximizing $r(x,\pi_1, \ldots, \pi_m)$ over $x \in \mathcal{R}$ for all $\pi_j \in \Pi$, $j\in J$.
This is true because for each item $j$ when there is a tie (Step 2 of Algorithm \ref{alg-round}) assigning the item to the bin with the highest value for the item, obviously produces a higher value.
To the best our knowledge, this is the first time that such an observation has been used for maximizing over a range.

\section{When Items are Held by Bidders} \label{sec:item-bidders}
Here, we mention the required modifications for the case where items are held by bidders rather than the bins.
Such bidders are called unit-demand bidders.
To better expose the changes we index bidders by $j$ in the following.
The allocation rule $\mathcal{A}$ takes reported valuations $v=(v_1,\ldots,v_m)$, $v_j=(v_{ij})_{i \in I}$ for all $j\in J$ and 
returns $(i_1,i_2,\ldots,i_m)$ where $i_j$ is the bin to which item $j$ is assigned.
Let $p_j$ denote the payment rule function for bidder $j$.
We use $v_j(i_1,\ldots,i_j,\ldots,i_m)$ instead of $v_{ij}$ for the sake of simplicity in the definition below.

\begin{definition}[truthful-in-expectation] \label{def-tie-2}
A mechanism is truthful-in-expectation for GAP (when items are held by bidders) if, for every bidder $j$, (true) valuation function $v_j$ , (reported) valuation
function $v'_j$, and (reported) valuation functions $v_{-j}$ of the other bidders,
\begin{equation} \label{eq-eps-tie-2}
\E[v_j(\mathcal{A}(v_j,v_{-j}))-p_j(v_j,v_{-j})] \geq \E[v_j(\mathcal{A}(v'_j,v_{-j}))-p_j(v'_j,v_{-j})] .
\end{equation}
The expectation in (\ref{eq-eps-tie-2}) is taken over the coin flips of the mechanism.
\end{definition}

For this type of bidders, we maximize over the range defined in Section \ref{sec:truthful} to obtain a MIDR. 
That is, we use Algorithm \ref{alg-gap-midr} as the allocation rule.
We need however a different payment rule for this type of bidders.

In order to calculate VCG fractional payment $p_j^{\textit{frac}}$, we need to calculate Clarke pivot $h_j(v_{-j})$ and the value gained by bidders other than $j$ in the current fractional solution, $F_{-j}(y^*)$.
We calculate $h_j(v_{-j})$ by running Algorithm \ref{alg-frac-local-search} after evaluating $v_{ij}$ to $0$ for all $i \in I$. 
Also, we have $F_{-j}(y^*)=\sum_{l=1, l\neq j}^m \sum_{i=1}^n \big(v_{\sigma_l(i),l}-v_{\sigma_l(i+1),l}\big) \big(1-\exp({-\sum_{k=1}^i y^*_{\sigma_l(k),l}})\big) $. 
By letting $w_j^{\textit{frac}}=F(y^*)-F_{-j}(y^*)$ and assuming that item $j$ is assigned to bin $i$ in the rounded solution, the payment of bidder $j$ is calculated as follows.

\begin{displaymath}
  \begin{array}{l l}
 p_j = \left\{ 
  \begin{array}{l l}
    \frac{v_{ij}}{w_j^{\textit{frac}}} p_j^{\textit{frac}} & \quad \text{if $w_j^{\textit{frac}} > 0$,}\\
    0 & \quad \text{if $w_j^{\textit{frac}} = 0$}.
  \end{array} \right. &
  \end{array}
\end{displaymath}

\section{Conclusion} \label{sec:conclusion}
We studied the problem of mechanism design for a strategic variant of GAP where valuations are assumed to be private information known only to the bidders while weights and capacities are publicly known.
Given that GAP is NP-hard, and that VCG is not trivially truthful with suboptimal solutions, we resorted to approximation mechanisms.

We proposed a solution by which the two obstacles, maximizing the social welfare of GAP as well as extracting true valuations of bidders, are surmounted. The solution provides bidders with incentives to report their valuations truthfully and runs in polynomial time approximating the social welfare with a provable ratio of at least $1-1/e$.

In comparison to the approximation algorithms presented for GAP without incentive issues, our proposed algorithm has advantages in terms of runtime and simplicity while presenting the same approximation ratio.
Our work also shows that the convex rounding technique is a powerful machinery for designing truthful approximation mechanisms and might find other applications in the field.

A problem which remains to be solved is the analysis of the strategic version of GAP in which weights and capacities are also private.
We conjecture there is no constant ratio truthful mechanism for this problem.

Another problem to be solved is to find a truthful mechanism for GAP with private values which stipulates that no item in the final allocation may remain unassigned.


\section*{Acknowledgments}\label{sec:Acknowledgments}

The authors gratefully acknowledge support of the Deutsche
Forschungsgemeinschaft (DFG) (BI 1057/7-1), and of the TUM Institute for
Andvanced Studies.

 \bibliographystyle{ACM-Reference-Format-Journals} 

 \bibliography{literature}

\end{document}